\newif\ifcheckpagelimits
\checkpagelimitstrue
\checkpagelimitsfalse

\ifcheckpagelimits
 \documentclass[nofootinbib,pra,aps,twocolumn,showpacs,showkeys,%
 amsmath,amssymb,superscriptaddress,final,reprint,floatfix,longbibliography]{revtex4-1}
 \newcommand{\todo}[1]{}
\else
 \documentclass[pra,aps,twocolumn,showpacs,showkeys,%
 amsmath,amssymb,superscriptaddress,final,reprint,floatfix,longbibliography]{revtex4-1}
 \newcommand{\todo}[1]{{\pdfmargincomment[icon=Note,color=pink]{#1}}}
\fi

\usepackage{lineno}
  \usepackage{mathptmx}
\usepackage{subfigure}
\usepackage{dcolumn}
\usepackage{amsmath,amssymb}
\usepackage{bm}
\usepackage{color}
\usepackage{overpic}
\usepackage{latexsym}
\usepackage{epstopdf}
\usepackage{color}
\usepackage[english]{babel}
\usepackage{latexsym}
\usepackage{stmaryrd}

\usepackage{braket}

\definecolor{mygrey}{gray}{0.35}
\definecolor{myblue}{rgb}{0.2,0.2,0.8}
\definecolor{myzard}{cmyk}{0,0,0.05,0}
\definecolor{mywhite}{rgb}{1,1,1}
\definecolor{myred}{rgb}{1,0.,0.3}

\usepackage[colorlinks=true,citecolor=myblue,linkcolor=myred]{hyperref}

 \def\ee{\mathord{\rm e}}
 
 \def\ii{\mathord{\rm i}}

\def\half{\textstyle\frac{1}{2}}

\renewcommand{\ii}{{\rm i}}
\renewcommand{\ee}{{\rm e}}
\def\beq{\begin{equation}}
\def\eeq{\end{equation}}

\begin{document}

\title{Dynamical polaron ansatz: a  theoretical tool for the ultra-strong coupling regime of circuit QED}

\author{Guillermo D{\'\i}az-Camacho}
\affiliation{Instituto de F{\'i}sica Fundamental IFF-CSIC, Calle Serrano 113b, Madrid 28006, Spain}
\author{Alejandro Berm{u}dez}
\affiliation{Instituto de F{\'i}sica Fundamental IFF-CSIC, Calle Serrano 113b, Madrid 28006, Spain}
\author{Juan Jos{\'e} Garc{\'\i}a-Ripoll}
\affiliation{Instituto de F{\'i}sica Fundamental IFF-CSIC, Calle Serrano 113b, Madrid 28006, Spain}

\begin{abstract}
In this work we develop a semi-analytical variational ansatz to  study  the properties of few photon excitations interacting with a collection of quantum emitters in  regimes that go beyond the rotating wave approximation. This method can be used to approximate both the static and dynamical properties of a superconducting qubit in an open transmission line, including the spontaneous emission spectrum and the  resonances in scattering experiments. The approximations are quantitatively accurate for rather strong couplings, as shown by a direct comparison to Matrix-Product-State numerical methods, and provide also a good qualitative description for stronger couplings well beyond the Markovian regime.
\end{abstract}
\maketitle

\setcounter{secnumdepth}{5}
\setcounter{tocdepth}{5}
\begingroup
\hypersetup{linkcolor=black}
\tableofcontents
\endgroup

\section{\bf Introduction}

The impressive progress in coupling single photons to single emitters,  both in the microwave\ \cite{astafiev} and in the optical domains\ \cite{lodahl}, allows us to talk about an emerging field of propagating-photon quantum technology. This field already demonstrates new devices that foresee interesting applications, such as few-photon transistors\ \cite{hoi11,chang07}, non-classical states of the radiation field\ \cite{hoi12,eichler12}, or photon-mediated interactions\ \cite{loo13,lalumiere12}.

The development of these technologies has been accompanied by numerous analytical and numerical techniques to model light-matter and light-mediated matter interactions. These include single-photon single-qubit effective boundary conditions\ \cite{shen05a,shen05b}, input-output theory\ \cite{shi11,peropadre13a} or scattering theory\ \cite{shi09,laakso14,shi15}. Nonetheless, the degree of control of some methods, or their computational generality for arbitrary numbers of photons, remains an open problem. Moreover, all these methods are restricted to the Rotating Wave Approximation (RWA) regime, in which counterrotating terms are neglected. The RWA breaks down, for instance, when the spontaneous emission rate of a two-level quantum emitter, $\gamma$, becomes comparable to its energy gap, $\Delta\sim \gamma$. Such deep ultra-strong coupling is now within experimental reach in the superconducting world\ \cite{ultrastrong_coupling,ultrastrong_couplingb,bourassa09,peropadre13}, and we expect that it will become also feasible in the near future in experiments with single emitters in contact with optical photons and plasmons.

The microscopic Hamiltonian describing light-matter interactions can be formally mapped, in any coupling regime, onto the so-called spin-boson model~\cite{sbm_rmp}. Hence, in dealing with strong and ultra-strong interactions, Optics can look into Condensed-Matter Physics for inspiration, generalising methods that work with the spin-boson Hamiltonians, and adapting them in order to  describe few propagating photonic excitations. Some  of these methods, such as the non-interacting-blip approximation (NIBA)\ \cite{sbm_rmp}, or Wilson's numerical renormalisation group (NRG)~\cite{nrg_spin_boson}, focus directly on the dynamics of the emitters, such that one cannot address the scattering of photons. Other methods, such as the sophisticated numerical techniques based on the Density-Matrix Renormalisation Group (DMRG) or Matrix-Product States (MPS), have a greater potential in dealing with these effects~\cite{peropadre13,sanchez-burillo14a,sanchez-burillo14b}, at the expense of a higher computational cost.

In this work, we study a third family of methods based on the Lang-Firsov transformation for the polaron problem\ \cite{lang_firsov}. These methods consist on a variationally-optimised unitary transformation that displaces the electromagnetic field based on the state of the two-level emitter, and has already provided valuable results for the equilibrium properties of the spin-boson model\ \cite{variational_ansatz_luther_emery,variational_ansatz_zwerger,variational_ansatz_Silbey_Harris}. We introduce a {\it dynamical polaron ansatz}, which is a time-dependent variational wavefunction describing the two-level emitter at zero temperature, together with the scattering states of the photons. We thus upgrade this family of variational methods to address relevant non-equilibrium problems, such as the spontaneous emission  or the ultra-strong  spectroscopy of the emitter. Additionally, these dynamical ansatzs lead to analytical results in certain regimes, which are valuable to develop a physical intuition about these complex non-equilibrium effects, and can be straightforwardly generalised to more complex situations with multiple quantum emitters coupled to propagating photons. The numerical and analytical methods are simpler than full MPS simulations, and a comparison with these shows good qualitative and even quantitative agreement up to very large coupling strengths, $\gamma\sim 0.4\Delta$ in the Ohmic spin-boson model. This shows the potential of these conceptually and technically simpler tools for analysing and designing future experiments in propagating-photon quantum technologies.

The outline of this work is as follows. In Sect.\ \ref{sec:spin-boson}, we introduce the spin-boson model as we use it to describe light-matter interaction of few two-level emitters, also refereed to as qubits, with a low-dimensional photonic waveguide or microwave transmission line. In Sect.\ \ref{sec:ansatzs}, we describe different variational ans\"atze. We start with two variational ans\"atze based on the Lang-Firsov transformation describing a static and a time-dependent wavefunction with up to one hybrid excitation with variational weights in the qubits or in the photonic modes. We also describe briefly the MPS methods that we compare with. In Sect.\ \ref{sec:numerics}, we apply all these methods to the study of a single quantum emitter in the photonic line, and its  interaction with the propagating photons. We show how the ansatz properly describes the qubit polarization in the open field, the frequency renormalization, and also the transmission and reflection coefficients for an incoming low-intensity photonic wave-packet with a broad distribution among the possible photonic modes, even in the ultra-strong coupling regime. Finally, in Sect.\ \ref{sec:conclusions}, we discuss our conclusions and possible extensions to treat multiple quantum emitters.

\section{\bf Spin-boson model}
\label{sec:spin-boson}
  
The  Hamiltonian describing a collection of quantum emitters interacting with a one-dimensional electromagnetic (EM) field in the  ultra-strong coupling regime corresponds to the well-known {\it few-impurity spin-boson model}\ \cite{sbm_rmp, vojta_review}, unitarily transformed to  a  rotated spin basis, namely
\begin{equation}
  \label{eq:spin-boson}
  H = \sum_{i} \frac{\Delta_i}{2}\sigma^z_i + \sum_k \omega_k a^\dagger_k a_k^{\phantom{\dagger}}
  + \sum_{ik} \sigma^x_i\left(g_{ik} a_k^\dagger + \mathrm{H.c.}\right).
\end{equation}
Here,  $a^\dagger_k, a_k^{\phantom{\dagger}}$ are  bosonic operators  that create and annihilate a quantum of the EM field with frequency $\omega_k$ labelled by  $k$ (e.g. in three-dimensions, $k$ contains the photon wavevector and its polarisation). The emitters are modelled as two-level atoms, also  referred to as qubits or spins,  $\sigma^z_i=\ket{{\uparrow_i}}\bra{{\uparrow_i}}-\ket{{\downarrow_i}}\bra{{\downarrow_i}}$, $\sigma^+_i=\ket{{\uparrow_i}}\bra{{\downarrow_i}}$,  $\sigma^-_i=\ket{{\downarrow_i}}\bra{{\uparrow_i}}$, $\sigma_i^x=\sigma_i^++\sigma_i^-$, and we have introduced  the transition frequency $\Delta_i$.
 
In the above Hamiltonian, the atom-photon interaction   is defined in terms of  the  couplings  
$ g_{ik} $, which  depend on the positions of the qubits $\boldsymbol{x}_i$. As customary in the quantum theory of radiation\ \cite{cohen_tannoudji_book}, the EM field acts as a bosonic reservoir that modifies the  dynamics of the quantum emitters depending on  the distribution of the  atom-photon couplings for  different frequencies, namely the spectral density 
\begin{equation}
\label{spectral_density}
  J_i(\omega) = 2\pi \sum_k |g_{ik}|^2\delta(\omega-\omega_k).
\end{equation}
For instance, when the atom-photon couplings are weak, and the  correlations of the EM bath decay sufficiently fast, a single excited qubit in contact with the EM field will decay exponentially with a rate $\gamma_i=J_i(\Delta)$\ \cite{weisskopf_wigner}, while a pair of qubits will show collective effects due to the exchange\ \cite{Fermi_problem}, or the spontaneous emission\ \cite{dicke_superradiance},  of EM photons. 

We shall consider both discrete and continuous descriptions of a one-dimensional EM reservoir, which model the photons in a transmission line.

\textit{(i) Discretized spin-boson model.-} A  transmission line of  length $L$ can be divided into $N$ segments of  length $\delta x=L/N$, which leads to  a discretized momentum  $k_n=\frac{2\pi}{L}n$, with $n\in\{0,\pm 1,\dots,\pm N/2\}$. The boson frequencies $\omega_k\to \omega_{k_n}$, which arise due to the couplings between the transmission line segments, and the  spin-boson couplings   $g_{ik}\to g_{ik_n}$, correspond to
 \begin{equation}
 \omega_{k_n}=\omega_{\rm c}\sqrt{2-2\cos\left(\frac{2\pi n}{N} \right)}, \hspace{2ex}
g_{ik_n} = g \sqrt{\frac{\omega_{k_n}}{2L}} \ee^{\ii \frac{2\pi n}{L} x_i}.
  \label{couplings_discrete}
 \end{equation}
 Here, we have introduced the cutoff frequency $\omega_{\rm c}=v/\delta x$, which  increases as the number of segments is raised. In the continuum limit $N\to\infty$,  and for energies well below the cutoff, one recovers the linear dispersion  $\omega_k\sim v|k|$ with $v$ playing the role of the speed of light in the transmission line. Let us note that this discretized model will be used  to perform numerical calculations. 

\textit{(ii) Continuum spin-boson model.-} Alternatively, we can model the transmission line directly in the  continuum limit $\delta x\to0$, where $k\in\mathbb{R}$, and the high-frequencies are exponentially  cut off  by substituting 
\begin{equation}
\omega_k=v|k|, \hspace{2ex}
g_{ik} = g \ee^{-\frac{\omega_k}{2\omega_{\rm c}}} \sqrt{\frac{\omega_k}{2L}} \ee^{\ii kx_i}.
\label{couplings_continuum}
\end{equation}
Let us note that the cutoff in this model can be set to any desired value, and the dispersion is assumed to be linear for all frequencies. This continuous model will be exploited to derive a number of analytical predictions. 
  
Both models of the bosonic bath lead to an Ohmic spectral density  at low-enough frequencies $\omega\ll\omega_{\rm c}$, which is identical for all qubits, namely
\begin{equation}
\label{ohmic}
J(\omega):=J_i(\omega)\approx \pi\alpha\omega,\hspace{1.5ex} \alpha=\frac{|g|^2}{\pi v},
\end{equation}
where $\alpha$ is a dimensionless spin-boson coupling strength that plays an important role. In the context of the  single-impurity spin-boson model, ultra-strong couplings 
lead to the so-called localization-delocalization  quantum phase transition $\alpha=1$\ \cite{loc_trans_chakravarty, loc_trans_bray}, and the coherent-incoherent dynamical crossover at $\alpha=1/2$\ \cite{chak_leggett,guinea_bosonization}.

As mentioned above, we shall explore the consequences of such effects in the static and dynamical properties for few quantum emitters~\eqref{eq:spin-boson} by using two types of variational ansatz\"{e}:

\textit{(i) Polaron variational ansatz.-} This ansatz was originally developed to understand the groundstate properties of the Kondo effect through its connection to the spin-boson model\ \cite{variational_ansatz_luther_emery,variational_ansatz_zwerger}, and has the non-variational Lang-Firsov transformation as its precursor\ \cite{lang_firsov}. The polaron ansatz builds on a variational family of spin-dependent coherent states, and correctly captures the quantum phase transition at $\alpha=1$. As shown in\ \cite{variational_ansatz_Silbey_Harris}, the polaron method also agrees with some of the finite-temperature  properties obtained by other methods\ \cite{sbm_rmp}, and can be combined with Markovian master equations to study the spin dynamics, although the predicted coherent-incoherent crossover does not coincide with $\alpha=1/2$. In this work, we introduce a \textit{dynamical polaron ansatz} that will allow us to overcome these limitations, and to develop a simple analytical understanding of how the archetypical quantum-optical properties of~\eqref{eq:spin-boson} predicted within the RWA \cite{weisskopf_wigner,Fermi_problem,dicke_superradiance}, are modified in the ultra-strong coupling regime. Moreover, this ansatz will allow us to address the dynamics of the propagating photons, which is extremely important given their experimental accessibility.
 
\textit{(ii) Matrix-Product-state variational ansatz.-} Introduced in Ref.\ \cite{finite_corr_states} and connected to DMRG\ \cite{white_dmrg} for one-dimensional quantum systems in\ \cite{ostlund_rommer}, MPS ans\"{a}tze have the non-variational valence-bond states as their precursor\ \cite{AKLT}. The MPS ansatz is defined as the product of a set of matrices, whose number is determined by the structure of the problem, and its size determines the  accuracy of the variational procedure. As explained below, in this work we use static and time-dependent MPS to study the spin-boson model in frequency space, generalising static and time-dependent techniques introduced in\ \cite{sbm_sc}.

The comparison between the quasi-exact numerical ansatz of MPS and the dynamical polaron ansatz applied to Eq.~\eqref{eq:spin-boson}, will allow us to assert the regimes of validity of the later. In our effort to make the polaron ansatz more familiar to the quantum optics community interested in going beyond the weak-coupling RWA regime, we shall determine the regimes of validity of its simpler analytical predictions.

\section{\bf Variational methods for the ultra-strong coupling regime}
\label{sec:ansatzs}

In this section, we describe  in detail the above variational techniques that can be used to study both static and dynamical effects for a collection of quantum emitters interacting with a 1D EM field through the spin-boson model~\eqref{eq:spin-boson}. We will describe the general approach, and leave its application to particular settings for the following sections. 
 
\subsection{Static polaron  ansatz}

The static polaron ansatz is a  variational method to approximate the equilibrium properties of a single-impurity spin-boson model, which has been applied to the Ohmic\ \cite{variational_ansatz_luther_emery,variational_ansatz_zwerger,variational_ansatz_Silbey_Harris} and sub-Ohmic\ \cite{SH_subOhmic} cases with significant success, specially in light of  its considerable simplicity. It can be  improved by enlarging the number of variational parameters\ \cite{multipolaron}, and generalised to the two-impurity spin-boson model\ \cite{two_spin_SH}. 

The static ansatz can be defined in terms of a variational polaron transformation, which captures the relevant correlations between the spins and the bosons, and acts on a product state where the spins and the bosons are not entangled. For an arbitrary number of spins $N_{\rm s}$, it can be defined as follows
\begin{equation}
\label{sh_ansatz}
\ket{\Psi_{\rm gs}^{\rm P}[f_{ik},c_{\boldsymbol{\sigma}}]}=U_{\rm P}^\dagger\left[f_{ik}\right]\ket{0}\otimes\ket{\psi_{\rm s}[c_{\boldsymbol{\sigma}}]}.
\end{equation}
Here, we have introduced the polaron unitary transformation
\begin{equation}
\label{pol_transf}
U_{\rm P}[f_{ik}]=\bigotimes_{i,k}\ee^{\sigma_i^x\left(f^*_{ik}a_k^{\dagger}-{\rm H.c.}\right)},
\end{equation} 
where $[f_{ik}]$ is the set of all  variational polaron parameters $f_{ik}\in\mathbb{C}$. The static polaron ansatz~\eqref{sh_ansatz} is defined in terms of the global bosonic vacuum   $\ket{0}$, and the variational  spin state 
\begin{equation}
\label{variational_spin_state}
\ket{\psi_{\rm s}[c_{\boldsymbol{\sigma}}]}=\hspace{-2ex}\sum_{\boldsymbol{\sigma}\in\{\uparrow,\downarrow\}^{ N_{\rm s}}}\hspace{-2ex}c_{\boldsymbol{\sigma}}\ket{{\sigma}_1}\otimes\ket{{\sigma}_2}\otimes\cdots\otimes\ket{\sigma_{N_{\rm s}}},
\end{equation}
which  depends on the  set $[c_{\boldsymbol{\sigma}}]$ of all variational spin parameters $c_{\boldsymbol{\sigma}}=c_{\sigma_1,\sigma_2,\cdots,\sigma_{N_{\rm s}}}\in\mathbb{C}$, fulfilling  $\sum_{\boldsymbol{\sigma}}|c_{\boldsymbol{\sigma}}|^2=1$.

The variational minimization over the spin-boson Hamiltonian~\eqref{eq:spin-boson},  defined as 
\begin{equation}
\epsilon_{\rm gs}^{\rm P}={\rm min}_{[f_{ik},c_{\boldsymbol{\sigma}}]}\{\bra{\Psi_{\rm gs}^{\rm P}[f_{ik},c_{\boldsymbol{\sigma}}]}H\ket{\Psi_{\rm gs}^{\rm P}[f_{ik},c_{\boldsymbol{\sigma}}]}\},
\end{equation}
can be expressed in terms of a   simpler minimization, 
\begin{equation}
 \epsilon_{\rm gs}^{\rm P}={\rm min}_{[f_{ik},c_{\boldsymbol{\sigma}}]}\{\bra{\psi_{\rm s}[c_{\boldsymbol{\sigma}}]}H_{\rm s}[f_{ik}]\ket{\psi_{\rm s}[c_{\boldsymbol{\sigma}}]}\},
\end{equation}
which requires diagonalising the following spin Hamiltonian instead of the original spin-boson model
\begin{equation}
\label{Ising_model}
H_{\rm s}[f_{ik}]=\sum_i\frac{\Delta_i}{2}\ee^{-\Xi_i[f_{ik}]}\sigma_i^z+\sum_{i,j}J_{ij}[f_{ik}]\sigma_i^x\sigma_j^x.
\end{equation}
This spin Hamiltonian   corresponds to  a long-range version of the paradigmatic Ising model in a transverse field\ \cite{itf}. This model displays  qubit frequencies that get exponentially  renormalized through
\begin{equation}
\label{pol_p}
\Xi_i[f_{ik}]=\sum_k f_{ik}^{\phantom{*}}f^*_{ik}+{\rm c.c.},
\end{equation}
and  photon-mediated Ising interactions with  strengths
\begin{equation}
\label{ising_p}
J_{ij}[f_{ik}]=\sum_k\left(\omega_kf_{ik}^{\phantom{*}}f_{jk}^*-g_{ik}^{\phantom{*}}f_{jk}^*-g_{ik}^*f_{jk}^{\phantom{*}}\right).
\end{equation}

Since the variational energy is a quadratic functional of the spin parameters, their minimization  is simple and can be carried out analytically for the single- or the two-impurity problem, which fixes  the spin parameters  in terms of the polaron ones. Accordingly, the  problem reduces to function minimization, and yields the optimal parameters $[f_{ik}^{\star}, c_{\boldsymbol{\sigma}}^{\star}]$, which are denoted with a star super-index, and shall be used in the following sections.  For more than two impurities, since the Ising interactions can have any particular pattern, an analytical solution cannot be obtained in general, and one must resort to Lanczos methods to extract the groundstate of the spin Hamiltonian, which can be efficiently implemented for reasonably high $N_{\rm s}$.

\subsection{Dynamical polaron ansatz}

Although understanding the static properties of the spin-boson model is already a non-trivial  problem, specially for a few impurities, a considerably more challenging task is to develop an accurate description of  non-equilibrium effects, and a number of techniques have been put forth over the years for that purpose\ \cite{sbm_rmp}.  As already mentioned in the introduction, most of these techniques can only address  qubit obervables, which is consistent with situations where the bosonic bath cannot be measured. However, with the advent of the new propagating-photon quantum technologies, this situation has been reversed, as the photonic properties of the setup are now accessible. The goal of this section is to introduce an accurate, yet simple, variational  ansatz that captures these dynamical effects for both the emitters and the photons.

In order to introduce such a dynamical polaron ansatz, let us revisit  the  spin Hamiltonian in Eq.~\eqref{Ising_model}, specified for the  optimal parameters $[f_{ik}^{\star}, c_{\boldsymbol{\sigma}}^{\star}]$ obtained with the static ansatz. The eigenstates of this  Hamiltonian contain, in addition to the variational groundstate $\ket{\psi_{\rm gs}}:=\ket{\psi_{\rm s}[c^{\star}_{\boldsymbol{\sigma}}]}$,  a number $N_{\rm e}$ of spin excitations $\{\ket{\psi_{\rm e}^{ s}}\}$ with energies $\{\epsilon_{\rm e}^s\}$ that can be excited if the qubits absorb a  photon from the EM environment. Inspired by our previous works on different quantum many-body models\ \cite{interspersed_sblm}, we can define a  dynamical variational ansatz by creating such  spin and photonic excitations over the polaron transformed groundstate, namely
\begin{equation}
\label{sh_ansatz_exc}
\ket{\Psi_{\rm exc}^{\rm P}[\alpha_s(t),\alpha_k(t)]}=U_{\rm P}^\dagger[f_{ik}^\star]W_{\rm sp}[\alpha_s(t),\alpha_k(t)]\ket{0}\otimes\ket{\psi_{\rm gs}},
\end{equation}
where we have introduced an operator that creates the relevant spin-photon excitation
\begin{equation}
W_{\rm sp}[\alpha_s(t),\alpha_k(t)]=\sum_{s=1}^{N_{\rm e}}\alpha_s(t)\ket{\psi^{s}_{\rm e}}\bra{\psi_{\rm gs}^{\phantom{+}}}+\sum_k\alpha_k(t) a_k^\dagger,
\end{equation}
with a certain set $[\alpha_s(t),\alpha_k(t)]$ of time-dependent variational parameters  $\alpha_s(t)\in\mathbb{C},\alpha_k(t)\in\mathbb{C}$.

By rearranging these  parameters in a complex-valued  vector  $\boldsymbol{\alpha}(t)=(\alpha_{1}(t),\cdots,\alpha_{{N_{\rm e}}}(t),\alpha_{k_1}(t),\alpha_{k_2}(t),\cdots )^{\rm t}$  fulfilling $\boldsymbol{\alpha}^\dagger(t)\boldsymbol{\alpha}(t)=1$, we can construct a  Lagrangian that leads to a time-dependent variational principle\ \cite{tdvp}, namely
\begin{equation}
\mathcal{L}_{\rm P}[\boldsymbol{\alpha}^\dagger,\boldsymbol{\alpha}]=\frac{\ii}{2}\left(\boldsymbol{\alpha}^\dagger\partial_t\boldsymbol{\alpha}- \partial_t\boldsymbol{\alpha}^\dagger\boldsymbol{\alpha}\right)-\mathcal{E}_{\rm P}[\boldsymbol{\alpha}^\dagger,\boldsymbol{\alpha}],
\end{equation}
where we do not write the explicit time-dependence to ease notation $\boldsymbol{\alpha}=\boldsymbol{\alpha}(t)$. Here, we have introduced the energy functional associated to the spin-boson Hamiltonian~\eqref{eq:spin-boson}, namely
\begin{equation}
\mathcal{E}_{\rm P}[\boldsymbol{\alpha}^\dagger,\boldsymbol{\alpha}]=\bra{\Psi_{\rm exc}^{\rm P}[\boldsymbol{\alpha}]}(H-\epsilon_{\rm gs}^{\rm P})\ket{\Psi_{\rm exc}^{\rm P}[\boldsymbol{\alpha}]},
\end{equation}
which is a quadratic functional of the variational parameters. Building a variational action from the above Lagrangian,  the  principle of minimal action\ \cite{tdvp} leads to a system of Euler-Lagrange equations  that describe the dynamics of the system restricted to the region of the Hilbert space spanned by the states parametrised by Eq.~\eqref{sh_ansatz_exc}. The accuracy of this variational method thus depends on our physically-motivated choice of the dynamical ansatz~\eqref{sh_ansatz_exc}, and shall be benchmarked by comparing to well-known properties for qubit observables of the single-impurity spin-boson model\ \cite{sbm_rmp}, and to our results of photon scattering  using time-dependent MPS simulations. 

For the simple parametrisation~\eqref{sh_ansatz_exc},  the  Lagrangian leads to a linear system of first-order differential equations $\ii\partial_t\boldsymbol{\alpha}=\mathbb{H}_{\rm P}\boldsymbol{\alpha}$. Here, the matrix $\mathbb{H}_{\rm P}$ can be obtained by evaluating the matrix elements
\begin{equation}
\label{matrix_elements}
\bra{0}\otimes\bra{\psi_{\rm gs}}W^\dagger_{\rm sp}[\boldsymbol{\alpha}](H_{\rm P}[f_{ik}^{\star}]-\epsilon_{\rm gs}^{\rm P})W_{\rm sp}[\boldsymbol{\alpha}]\ket{0}\otimes\ket{\psi_{\rm gs}},
\end{equation}
for  the polaron-transformed spin-boson  Hamiltonian
\begin{align}
H_{\rm P}[f_{ik}^{\star}]&=\sum_i\frac{\Delta_{i}}{2}\left(\sigma_i^z\cos\Theta_i[f_{ik}^{\star}]-\sigma_i^y\sin\Theta_i[f_{ik}^{\star}]\right)
\nonumber\\
&+\sum_k\omega_k a_k^{{\dagger}}a_k^{\phantom{\dagger}}+\sum_{i,j}J_{ij}[f_{ik}^{\star}]\sigma_i^x\sigma_j^x\nonumber\\
&+\sum_{ik}\sigma_i^x\left(\big(g_{ik}-\omega_kf_{ik}^\star\big)a_k^{\phantom{\dagger}}+{\rm H.c.}\right),
\end{align}
which depends on the  operator 
\begin{equation}
\Theta_i[f_{ik}^{\star}]=-2\ii\sum_k(f_{ik}^\star a_k^{\phantom{\dagger}}-{\rm H.c.}).
\end{equation}

By diagonalizing the matrix $\mathbb{H}_{\rm P}$, one finds the excitation energies and eigenstates, which are an admixture of the spins and photons, and can be understood as some sort of spin-photon waves. However, if one is interested in  the reduced
dynamics of either the spins, as customary in studies of the spin-boson model\ \cite{sbm_rmp}, or an incoming photonic wave-packet for  transmission-reflection experiments, one may try to develop a Weisskopf-Wigner-type theory\ \cite{weisskopf_wigner} for the above system of first-order differential equations. These equations can always be rewritten as
\begin{align}
\label{WW_Nqubit}
\ii\partial_t\alpha_s(t)&=\Delta_{s}\alpha_s(t)+\sum_k\mathfrak{g}_{ks}\alpha_k(t),\\
\ii\partial_t\alpha_k(t)&=\omega_k\alpha_k(t)+\sum_s\mathfrak{g}_{ks}^*\alpha_s(t)+\sum_{k'}\mathfrak{f}_{kk'}^*\alpha_{k'}(t),
\nonumber
\end{align}
where we have introduced the  energy of the spin excitations $\Delta_s=\epsilon_{\rm e}^s-\epsilon_{\rm gs}^{\rm P}$, and the couplings $\mathfrak{g}_{ks}$, and $\mathfrak{f}_{kk'}$, which can be obtained from the matrix elements of  $\mathbb{H}_{\rm P}$. Note that the couplings fulfil $\mathfrak{f}_{kk'}=\mathfrak{f}_{k'k}$, and can be thus  diagonalized by an orthogonal transformation $\Delta\omega_k\delta_{kk'}=\sum_{k_1,k_1'}\mathcal{M}_{k_1k}\mathfrak{f}_{k_1k_1'}\mathcal{M}_{k_1'k'}$. Accordingly, we can transform the variational parameters $\tilde{\alpha}_k(t)=\sum_{k'}\mathcal{M}_{k'k}\alpha_{k'}(t)$, and the remaining couplings $\tilde{\mathfrak{g}}_{ks}=\sum_{k'}\mathcal{M}_{k'k}\mathfrak{g}_{k's}$, such that
\begin{align}
\label{WW_N_qubit_transformed}
\ii\partial_t\alpha_s(t)&=\Delta_{\rm r}\alpha_s(t)+\sum_k\tilde{\mathfrak{g}}_{ks}\tilde{\alpha}_k(t),\\
\ii\partial_t\tilde{\alpha}_k(t)&=(\omega_k+\Delta\omega_k)\tilde{\alpha}_k(t)+\sum_s\tilde{\mathfrak{g}}_{ks}^*\alpha_s(t).
\nonumber
\end{align}
This system of equations resembles the Weisskopf-Wigner equations of spontaneous emission of two-level atoms coupled to the EM field\ \cite{weisskopf_wigner}. Let us highlight, however, that they are valid beyond the RWA intrinsic to the standard Weisskopf-Wigner theory thanks to the polaronic variational methods.
If one is interested in the reduced dynamics of the quantum emitters, as usual in the spin-boson model or in the theory of quantum radiation, one can develop a Weisskopf-Wigner-type\ \cite{weisskopf_wigner}  theory by formally integrating the equations for the bosonic amplitudes, and substituting them in the equations for the qubit amplitudes. Conversely, one may be interested in the scattering of propagating photons from the collection of quantum emitters, which would require the opposite process. In the following sections, we shall use both approaches, highlighting the importance of taking into account non-Markovian effects in the ultra-strong coupling regime.

Let us  note at this point that if the initial state contains some atomic coherences, the ansatz~\eqref{sh_ansatz_exc} must be generalised to include  also an amplitude in the groundstate $W_{\rm sp}[\alpha_s(t),\alpha_k(t)]\to W_{\rm sp}[\alpha_s(t),\alpha_k(t)]+\alpha_{\rm gs}(t)$. However, this amplitude does not contribute with any term in the evaluation of Eq.~\eqref{matrix_elements}, and thus $\alpha_{\rm gs}(t)=\alpha_{\rm gs}(0)$, whereas the time-evolution of the remaining variational parameters is still described by Eq.~\eqref{WW_N_qubit_transformed}. Yet, including this groundstate amplitude may be necessary to calculate the dynamics of certain observables, such as the atomic coherences that are important for the coherent-incoherent transition of the spin-boson model.

At this stage, it is worth pointing out that our dynamical ansatz~\eqref{sh_ansatz_exc} differs from the  application  of the so-called  Davidov ansatzs, which arise in the study of  exciton-phonon interactions, to the spin-boson model\ \cite{davidov_ansatz}. There are  two crucial differences: {\it (i)} Our ansatz is built in two steps, such that the polaron parameters are already fixed during the computation of the dynamical properties. This contrasts the Davidov ansatzs, which consists on time-dependent polaron parameters with additional time-dependent variational weights for each spin state. {\it (ii)} Our ansatz considers also additional single-photon excitations, which are absent in the Davidov ansatzs\ \cite{davidov_ansatz}. Property {\it (i)} will be crucial to be able to derive analytical expressions for the dynamics, whereas property {\it (ii)} will be crucial to describe the effect of spontaneous emission and photon scattering.

\subsection{Matrix-Product-State  ansatz}

The previous ansatz\"{e} will be compared with a well-established method\ \cite{sbm_sc} for the numerical simulation of the spin-boson model, which combines ideas from the quantum impurity ansatz\ \cite{quantum_impurity}, matrix product operators\ \cite{mpos}, and Arnoldi-type evolution methods\ \cite{mps-t-evolve}. More precisely, we write down a variational wavefunction for the qubit-photon system as
\begin{equation}
  \ket{\psi_A} = \mathrm{tr}(A^{s_1}A^{n_1}\cdots A^{n_M})\ket{s_1,n_1\ldots n_M},
\end{equation}
where $s_1$ is the quantum state of the qubit, $n_i$ are the Fock states of $M$ different photon modes in frequency space, and the $A$'s are different matrices with a size of up to $\chi^2\times n_{max}$, where $\chi$ is the bond dimension of the MPS ansatz and $n_{max}$ is the maximum occupation of the bosonic modes. In addition to this encoding of the wavefunction, we efficiently write the spin-boson model as a long-range-interaction matrix product operator (MPO), which has a rather small bond dimension, $\mathcal{O}(3)$.

Combining the efficient representation of the Hamiltonian with the MPS wavefunction, we can either compute approximations to the ground state of spin-boson model, or implement a time-evolution algorithm. In the first case we work by minimizing the energy functional
\begin{equation}
  E[\{A\}] = \frac{\braket{\psi_A | H |\psi_A}}{\braket{\psi_A | \psi_A}},
\end{equation}
with respect to the collection of numbers in all the matrices/tensors $A$. The minimization procedure is efficient thanks to the MPO representation of the Hamiltonian, even when it contains long-range interactions, and it is implemented with a generalisation of the DMRG sweeping technique.

The time evolution is implemented using an approximation of the exponential of the Hamiltonian for short times. More precisely, we construct $\exp(-iH\Delta{t})\ket{\psi_A}$ as a linear combination of MPS, $\sum_n c_n \ket{\phi_n}$, where the vectors $\ket{\phi_n}$ form a Krylov basis built with MPS themselves, as explained in Ref.\ \cite{mps-t-evolve}. The use of Arnoldi expansions allows us to profit from the MPO expansion of the Hamiltonian and work with the long-range interactions, something which is much harder with Trotter expansions.

It has to be remarked that, while more accurate than the dynamical polaron ansatz, the MPS method is more complex from the implementation and computational point of view. The number of parameters in an MPS variational wavefunction scale as $N\times \chi^2 n_{max}$, while in our few photon dynamical ansatz with one qubit, we have at most $(N+1)\times 2$ degrees of freedom. It is therefore interesting to use the MPS as a benchmark with which to assert the range of validity of the polaron ansatz, with the idea of both having a flexible and simpler tool, and also a way to implement potential analytical approximations and effective models.

\section{\bf Single quantum emitter applications}
\label{sec:numerics}

Once the different variational ansatz\"{e} have been described, let us  apply them to the simplest possible scenario: a single quantum emitter ultra-strongly coupled to a 1D EM field. We will exploit  the analytic polaron predictions based on the continuum model~\eqref{couplings_continuum} to offer physical insight, and benchmark the numerical  polaron results based on the discretised model~\eqref{couplings_discrete} with MPS simulations for an identical  discretisation. The main objective is to prove that the simple polaron ansatz, in comparison to the  more involved MPS technique, provides a sufficiently accurate description of both static and dynamical phenomena, with the hope that  it will  be established as a simple theoretical tool within the quantum optics community dealing with the ultra-strong coupling regime. 

\subsection{Static predictions}

\subsubsection{Continuum spin-boson model}

Let us  consider the solution of the variational system of equations for the continuum single-impurity spin-boson model in Eqs.~\eqref{eq:spin-boson} and~\eqref{couplings_continuum}~\cite{variational_ansatz_luther_emery,variational_ansatz_zwerger,variational_ansatz_Silbey_Harris}. For  $N_{\rm s}=1$,  the  Ising Hamiltonian~\eqref{Ising_model} reduces to a single-spin problem, and the energy functional is  
\begin{equation}
 \epsilon_{\rm gs}^{\rm P}={\rm min}_{[f_{1k}]}\left\{J_{11}[f_{1k}]-\half\Delta_{{\rm r}}[f_{1k}]\right \},
 \end{equation}
  where we have introduced the renormalised frequency
 \begin{equation}
 \label{ren_freq}
\Delta_{{\rm r}}=\Delta_1\ee^{-\Xi_1[f_{ik}]}.
 \end{equation}
 The function minimisation yields
 \begin{equation}
 \label{eqs_f_single_qubit}
 f_{1k}=\frac{g_{1k}}{\omega_k+\Delta_{{\rm r}}[f_{1k}]},
 \end{equation}
 which amounts to a non-linear system of equations for the polaron parameters. In this case, the variational spin parameters are
  \begin{equation}
\label{eqs_c_single_qubit}
c_{\sigma_1}=\delta_{\sigma_1,\downarrow}\theta(\Delta_{{\rm r}}[f_{1k}]),
\end{equation}
where $\theta(x)$ is the Heaviside step function (i.e. if the renormalized qubit frequency vanishes, the variational groundstate corresponds to the two-fold degenerate Lang-Firsov transformed state).
Therefore, by solving the system of implicit equations~\eqref{eqs_f_single_qubit} and substituting in Eq.~\eqref{eqs_c_single_qubit}, we can recover the variational groundstate~\eqref{sh_ansatz} and calculate any observable. Let us note that by applying a non-variational perturbative approach in the polaron-transformed picture, the same condition~\eqref{eqs_f_single_qubit} has been found by imposing that the first-order perturbations vanish\ \cite{perturb_approach}, as customary in spin-wave approaches\ \cite{var_mf_xy}.

Note that  the system of equations~\eqref{eqs_f_single_qubit} can be rewritten in terms of a single implicit equation for the renormalized qubit frequency
\begin{equation}
\Delta^{\rm cp}_{{\rm r}}=\Delta\hspace{0.2ex}{\rm exp}\left\{-\int_{0}^\infty\hspace{-1.5ex}{\rm d}\omega \frac{J(\omega)}{\pi \left(\omega+\Delta_{{\rm r}}\right)^2}\right\},
\end{equation} 
where the super-index ${\rm ``cp"}$ stands for the continuum polaron model, and the exact spectral function corresponding to Eq.~\eqref{couplings_continuum} is $J(\omega)=\pi\alpha\omega\ee^{-\omega/\omega_{\rm c}}$. This equation can be solved analytically in the so-called scaling limit $\omega_{\rm c}\gg\Delta\geq\Delta_{\rm r}$, where one finds
\begin{equation}
\label{Deltar_continuum_single_qubit}
\Delta_{\rm r}^{\rm cp }=\Delta\left(\frac{p\Delta}{\omega_{\rm c}}\right)^{\frac{\alpha}{1-\alpha}},
\end{equation}
 with $p=\ee^{1+\gamma}$, and $\gamma$ as  Euler's constant. Therefore, the polaron ansatz predicts that the frequency of the quantum emitter gets renormalized as a consequence of its coupling to the photonic excitations (i.e. a photonic 'polaron' cloud dresses the quantum emitter and leads to a renormalized transition frequency). This  agrees with  adiabatic renormalisation group arguments\ \cite{sbm_rmp}, and  locates the localization-delocalization  transition at $\alpha=1$, where the qubit frequency vanishes. This is an example of the of the so-called boundary quantum phase transitions that also arise in other condensed-matter contexts\ \cite{vojta_review}.
 
 Using this result, we can recover the polaron~\eqref{eqs_f_single_qubit} and spin~\eqref{eqs_c_single_qubit} optimal parameters, which shall be denoted as   $[f_{1k}^{\star}, c_{\sigma_1}^{\star}]$,  and calculate any static observable by constructing the polaron  groundstate~\eqref{sh_ansatz}, such as 
\begin{equation}
\label{polarization_single_qubit}
\langle\sigma_1^x\rangle_{\rm gs}^{\rm cp}=\langle\sigma_1^y\rangle_{\rm gs}^{\rm cp}=0,\hspace{2ex}\langle\sigma_1^z\rangle_{\rm gs}^{\rm cp}=-\frac{\Delta_{\rm r}^{\rm cp}}{\Delta}.
\end{equation}
For weak spin-boson interactions $\alpha\ll1$, $\langle \sigma_1^z\rangle\to-1$ as expected from the groundstate of a bare qubit $\ket{{\downarrow}}$. However, as the coupling to the bosonic bath is raised, the groundstate starts populating also the excited bare state  $\ket{{\uparrow}}$, until $\langle \sigma_1^z\rangle\to0$ for $\alpha\to 1$. In the usual rotated basis of the spin-boson model\ \cite{sbm_rmp}, this corresponds to the localised phase.

\subsubsection{Discretised spin-boson model}

As emphasised previously, the continuum model serves to gain analytical insight into the static effects. However, it is the discrete version which can be benchmarked 
with  MPS simulations and, more importantly, the model that provides a more accurate microscopic description of the bosonic bath in the physical transmission line. The solution of the variational problem in this discretised spin-boson model requires the use of numerics, and has not been considered previously in the literature.

In the discretised model in Eqs.~\eqref{eq:spin-boson} and~\eqref{couplings_discrete},    the transmission line is divided into coupled segments, such that the cutoff frequency depends on the input parameters $L$ and $N$, being $L$ the total physical length of the line, and $N$ the number of segments of length $\delta x = L/N$ in which it is  divided. We are interested in raising the number of segments $N$ for a fixed length $L$, such that  the cutoff $\omega_c=v/\delta x$ can be maximised towards the above scaling limit,  and the linear region of the dispersion relation  $\omega_k\sim v|k|$ around the qubit frequencies $\Delta_i$ contains as many photonic modes as possible. However, we also note that the computational complexity of the polaron/MPS ansatz grows with $N$, as the number of bosonic modes also increases,   $k_n=\frac{2\pi}{L}n$ with $n\in\{0,\pm 1,\dots,\pm N/2\}$, and we cannot consider an arbitrarily-large $N$.

In the discretised polaron ansatz, we  numerically diagonalize  the transformed Hamiltonian~\eqref{Ising_model} to obtain the state of minimum energy as a function of the variational parameters $[f_{1k_n},c_{\sigma_1}]$. Then, we obtain  the optimal parameters $[f^\star_{1k_n},c_{\sigma_1}^\star]$ by using a numerical optimization routine, minimizing this energy with respect to the variational parameters. Once these optimal parameters are obtained, we can calculate numerically parameters such as the renormalized qubit frequency, taken directly from  the displaced Hamiltonian~\eqref{Ising_model}
\begin{equation}
\Delta_{\rm r}^{\rm dp}=\Delta \ee^{-2\sum_n |f^{\star}_{1k_n}|^2 },
\end{equation} 
where the index ``dp" stands for the discretised polaron.

\begin{figure}
\centering
\includegraphics[width=.85\columnwidth]{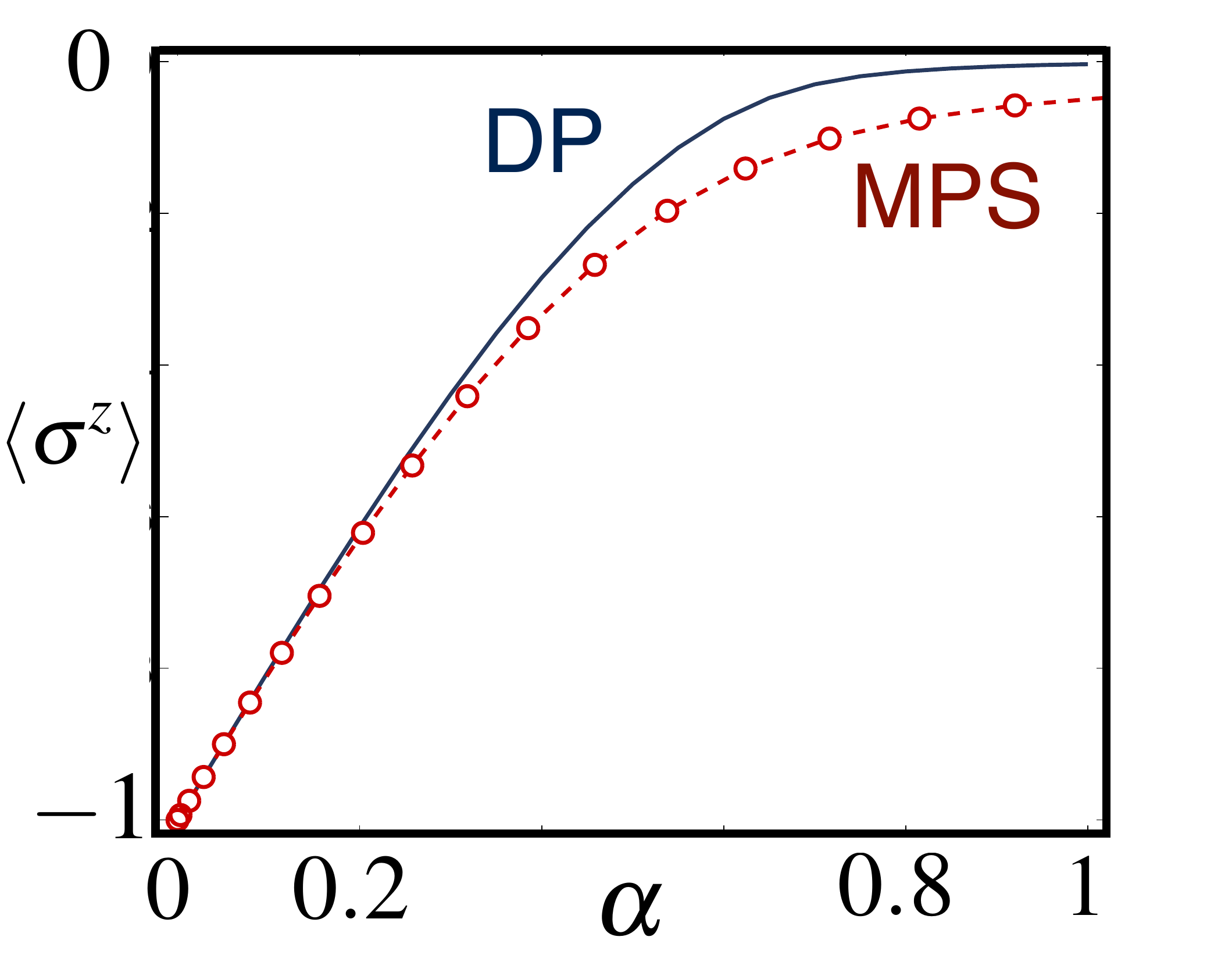} 
\caption{ {\bf Qubit polarization in the  single-impurity spin-boson model:} Groundstate  polarization  $\langle \sigma_1^z\rangle_{\rm gs}^{\eta}$ as a function of the spin-boson coupling $\alpha=|g|^2/\pi v$. In dotted red line, the value from MPS simulations $\langle \sigma_1^z\rangle_{\rm gs}^{\rm MPS}$ for $N=301$ and $L=10\lambda_0$. In blue solid line, we represent  $\langle \sigma_1^z\rangle_{\rm gs}^{\rm dp}$  for the discrete ansatz, with $L=10\lambda_0$ and $N=301$.}
\label{fig_sz_all}
\end{figure}

In order to benchmark  these static predictions with the MPS simulations, we should focus on some expectation values that can also be obtained through MPS. An observable of interest might be the qubit polarisation, whose calculation is simple using  the  polaron ground-state $ \langle \sigma_1^z \rangle_{\rm gs}^{\rm dp}=\bra{\psi_{\rm s}[c_{\sigma_1}^\star]}U_{\rm P}^{\dagger}[f_{1k_n}^\star]\sigma_1^z U_{\rm P}[f_{1k_n}^\star]\ket{\psi_{\rm s}[c_{\sigma_1}^\star]}$. After the numerical optimisation, which yields $\ket{\psi_{\rm s}[c_{\sigma_1}^\star]}$ for the discretised spin-boson model, we obtain the polarisation by calculating
\begin{equation}
 \langle \sigma_1^z \rangle_{\rm gs}^{\rm dp}=\bra{\psi_{\rm s}[c_{\sigma_1}^\star]}\sigma_1^z  \ket{\psi_{\rm s}[c_{\sigma_1}^\star]}e^{-2\sum_n |f^\star_{1k_n}|^2} .
\end{equation}
Note that the eigenvectors are displaced by the polaron transformation~\eqref{pol_transf}, but so does the spin operator (hence the exponential due to the renormalization), so the result will be on the correct frame of reference. We represent these results in Fig.~\ref{fig_sz_all} with the same results from the MPS simulations $ \langle \sigma_1^z \rangle_{\rm gs}^{\rm MPS}$, where we observe a reasonably-good agreement between the polaron and MPS results for the same discretisation with $N=301$ bosonic modes, which still lies far away from the scaling limit. We note that the discrete polaron results can be extended to a finer discretisation with $N=601$, and approach the prediction of the continuum polaron~\eqref{polarization_single_qubit} in the scaling limit. Such high number of bosonic modes compromises the accuracy of the MPS simulation, and highlights the ultimate power of the computationally less expensive polaron methods.

%%%%%%%%
%%%%%%%%%%

\subsection{Spontaneous emission}
	
So far, we have been concerned with the static properties of an ultra-strongly coupled qubit, although an even  richer phenomenology arises out of equilibrium. In the context of the quantum theory of radiation, the typical situation is to study the evolution of an initially excited quantum emitter, whose population decays irreversibly as a consequence of the photonic reservoir. By looking at the coherences, one may find the analogue of the coherent-incoherent dynamical crossover of the spin-boson model, where the spin displays a transition between damped oscillations and over-damped decay as the spin-boson coupling strength is increased beyond $\alpha=1/2$. The correct prediction of this dynamical effect is more challenging than the localization-delocalization transition, and requires  more involved techniques\ \cite{chak_leggett,guinea_bosonization} than the combination of the polaron static ansatz with a Markovian master equation\ \cite{variational_ansatz_Silbey_Harris,SH_subOhmic}. From  previous non-variational techniques\ \cite{convolution_meq}, it becomes clear that the Markovian assumption must be abandoned if one wants to capture the correct dynamical behaviour. 
	
The objective of this section is to study the evolution of the populations and coherences of an initially-excited quantum emitter, ultra-strongly coupled to the 1D EM field by applying the  dynamical ansatzs introduced  previously, and we will show that the above previous limitations can be overcome with our method.
	
\subsubsection{Continuum spin-boson model}

For the single-impurity case $N_{\rm s}=1$,  the situation simplifies as the  Ising Hamiltonian~\eqref{Ising_model} reduces to a single-spin problem, and we have a simple groundstate $\ket{\psi_{\rm gs}}=\ket{{\downarrow}}$ together with a single spin excitation $\ket{\psi_{\rm e}^1}=\ket{{\uparrow}}$,  provided that  $\Delta_{\rm r}^{\rm cp}>0$. We can thus easily build the dynamical ansatz~\eqref{sh_ansatz_exc}, and obtain the corresponding differential equations~\eqref{WW_Nqubit}, which are the variational analogue of the Weisskopf-Wigner theory\ \cite{weisskopf_wigner} of spontaneous emission
\begin{equation}
\label{WW_1qubit}
\begin{split}
\ii\partial_t\alpha_1(t)&=\Delta_{\rm r}^{\rm cp}\alpha_1(t)+\sum_k\mathfrak{g}_{k1}\alpha_k(t),\\
\ii\partial_t\alpha_k(t)&=\omega_k\alpha_k(t)+\mathfrak{g}_{k1}^*\alpha_1(t)+\sum_{k'}\mathfrak{f}_{kk'}^*\alpha_{k'}(t).
\end{split}
\end{equation}
Here, the following couplings between the spin and photonic excitations arise
\begin{equation}
\mathfrak{g}_{k1}=\frac{2\Delta^{\rm cp}_{\rm r}}{({\omega_k}+\Delta^{\rm cp}_{\rm r})}g_{1k},
\end{equation}
together with the additional couplings between the bare photonic modes
\begin{equation}
\label{mode_coupling_1qubit}
\mathfrak{f}_{kk'}=\frac{\Delta^{\rm cp}_{\rm r}}{(\omega_k+\Delta^{\rm cp}_{\rm r})(\omega_{k'}+\Delta^{\rm cp}_{\rm r})}\bigg(g_{1k}^{\phantom{*}}g_{1k'}^*+{\rm c.c.}\bigg),
\end{equation}
which are symmetric $\mathfrak{f}_{kk'}=\mathfrak{f}_{k'k}$, as announced in the section introducing the dynamical ansatz.  Hence, this coupling matrix  can be  diagonalized by an orthogonal transformation leading to the general equations~\eqref{WW_N_qubit_transformed}, which read in this case  
\begin{equation}
\label{WW_1_qubit}
\begin{split}
\ii\partial_t\alpha_1(t)&=\Delta_{\rm r}^{\rm cp}\alpha_1(t)+\sum_k\tilde{\mathfrak{g}}_{k1}\tilde{\alpha}_k(t),\\
\ii\partial_t\tilde{\alpha}_k(t)&=(\omega_k+\Delta\omega_k)\tilde{\alpha}_k(t)+\tilde{\mathfrak{g}}_{k1}^*\alpha_1(t).
\end{split}
\end{equation}

We  integrate out the photonic modes by first changing to a rotating frame $\alpha'_1(t)=\ee^{\ii\Delta^{\rm cp}_{\rm r}t}\alpha_1(t)$, $\tilde{\alpha}'_k(t)=\ee^{\ii(\omega_k+\Delta\omega_k)t}\tilde{\alpha}_k(t)$, and then substituting  the following expression 
\begin{equation}
\tilde{\alpha}'_k(t)=-\ii\int_0^t{\rm d}\tau\tilde{\mathfrak{g}}^*_{k1}\ee^{\ii(\omega_k+\Delta\omega_k-\Delta_{\rm r}^{\rm cp})\tau}\alpha'_1(\tau),
\end{equation}
where we have assumed that there are no  photonic excitations in the initially excited state, except for those intrinsic to the polaron cloud dressing the emitter. This leads to an integro-differential equation of the convolution type
\begin{equation}
\label{int_diff_eq}
\partial_t\alpha'_1(t)=-\int_0^t{\rm d}\tau K_1(t-\tau)\alpha_1'(\tau),
\end{equation}
where we have defined the following memory kernel 
\begin{equation}
K_{1}(t)=\sum_k|\mathfrak{g}_{k1}|^2\ee^{-\ii (\omega_k+\Delta\omega_k-\Delta_{\rm r}^{\rm cp})t}.
\end{equation}
Since we are interested in deriving some analytical formulas, we need to replace all the sums by integrals over the spectral density. To do so, we note that the frequency shifts of the photons $\Delta\omega_k$ contribute at a higher-order of the spin-boson coupling, and neglecting them yields
\begin{equation}
K_{1}(t)=\int_0^\infty\frac{J(\omega)}{2\pi}\left(\frac{2\Delta_{\rm r}^{\rm cp}}{\omega+\Delta_{\rm r}^{\rm cp}}\right)\ee^{-\ii (\omega-\Delta_{\rm r}^{\rm cp})t}.
\end{equation}

{\it (a) Markovian approximation:}  As customary in the Weisskopf-Wigner theory\ \cite{weisskopf_wigner}, we perform a change of variables $\tau'=t-\tau$ in the convolution~\eqref{int_diff_eq}, and a Markovian approximation to extend the integration domain to $\tau'\in\mathbb{R}^+$, and substitute $\alpha_1'(\tau)\to\alpha_1'(t)$ in Eq.~\eqref{int_diff_eq}.  In this case, 
 after using  $\int_0^\infty{\rm d}\tau'\ee^{-\ii\omega\tau'}=\pi\delta(\omega)-\ii\mathcal{P}(\omega^{-1})$ where $\mathcal{P}$ stands for Cauchy's principal value, the differential equation for the qubit can be expressed as
\begin{equation}
\label{markovian_WW_1}
\ii\partial_{t}\alpha'_1(t)=\left(\delta_1-\ii\frac{\gamma_1}{2}\right)\alpha'_1(t),
\end{equation}
where we have introduced the single-qubit decay rate $\gamma_1$,  as well as the  single-qubit  Lamb shift $\delta_1$. The decay rate within this Markovian approximation can be easily evaluated
\begin{equation}
\label{rate_cp}
\gamma_1=J(\Delta_{\rm r}^{\rm cp})=\pi\alpha\Delta\left(\frac{p\Delta}{\omega_{\rm c}}\right)^{\frac{\alpha}{1-\alpha}}.
\end{equation}
This yields a very sensible result: the decay rate, which is given by the value of the spectral function evaluated at the bare qubit frequency $\gamma_1^{\rm RWA}=J(\Delta)$  within the usual RWA and weak-coupling assumptions\ \cite{cohen_tannoudji_book},  must be substituted by the value of the spectral function at the renormalized qubit frequency~\eqref{Deltar_continuum_single_qubit} according to Eq.~\eqref{rate_cp}. In agreement with more involved methods\ \cite{sbm_rmp}, our simple dynamical polaron ansatz predicts that the localization-delocalization transition at $\alpha=1$, where the renormalized qubit frequency vanishes, is also accompanied by a vanishing decay rate $\gamma_1=0$.

In this Markovian regime, and in the scaling limit, the Lamb-type shift can be calculated after solving the Principal value integral, and yields
\begin{equation}
\label{lamb_cp}
\delta_1=-\alpha\Delta_{\rm r}^{\rm cp}.
\end{equation}
It is  interesting to note that, in analogy to the original calculation of the Lamb shift of the EM field where the self-energy is  subtracted, the variational polaron formalism includes  this self-energy  directly  in the groundstate energy instead of the Lamb shift. This contrasts the calculations within the RWA\ \cite{rwa_paper}, where the self-energy is not subtracted, and the Lamb shifts diverge with the cutoff frequency $\delta_i^{\rm RWA}=-\alpha(\omega_{\rm c}-\half\Delta\log(\Delta/\omega_{\rm c}))+\frac{\alpha}{2}\gamma\Delta$.

If we take the Lamb shift~\eqref{lamb_cp} together with the polaron renormalisation, the frequency of the quantum emitter becomes $\Delta_{\rm r}^{\rm cp}(1-\alpha)$, which predicts that the evolution of coherences will stop at $\alpha=1$, instead of the prediction $\alpha=1/2$ of the coherent-incoherent transition by other methods\ \cite{sbm_rmp}. As we argue below, this is an artefact of the Markovian approximation, and can be cured by a more careful analysis.

\vspace{0.5ex}

{\it (b) Non-Markovian approximation:} Let us reconsider the integro-differential equation using resolvent-operator techniques\ \cite{van_hove,davidovich,cohen_tannoudji_book}. By Laplace transform, and using the Bromwich contour to invert it, one can express the solution  as $\alpha'_1(t)=\frac{\ii}{2\pi}\int_{-\infty}^{\infty}{\rm d}\epsilon\ee^{-\ii\epsilon t}G(\epsilon)$, where we have introduced the following propagator
\begin{equation}
G(\epsilon)=\frac{1}{(\epsilon-\Delta_{\rm r}^{\rm cp}+\delta(\epsilon))+\ii\frac{\gamma(\epsilon)}{2}},
\end{equation}
and we have defined the so-called level-shift operator
\begin{equation}
\delta(\epsilon)=-\frac{2\alpha(\Delta_{\rm r}^{\rm cp})^2}{\Delta_{\rm r}^{\rm cp}+\epsilon},
\end{equation}
obtained from the principal value integral, and the so-called level broadening operator
\begin{equation}
\gamma(\epsilon)=J(\epsilon)\left(\frac{2\Delta_{\rm r}^{\rm cp}}{\epsilon+\Delta_{\rm r}^{\rm cp}}\right)^2.
\end{equation}
Note that, if one simply substitutes $\epsilon\to\Delta_{\rm r}^{\rm cp}$ in both operators, which is known as the single-pole approximation, one can easily perform the integral recovering the Markovian rate~\eqref{rate_cp} and Lamb-type shift~\eqref{lamb_cp}. However, if we are interested in the coherent-incoherent transition, a more careful analysis is required. In particular, by looking at solutions of $\epsilon-\Delta_{\rm r}^{\rm cp}+\delta(\epsilon)=0$ where the propagator gets maximal, one finds $\epsilon_m=\sqrt{1-2\alpha}\Delta_{\rm m}$. A Taylor expansion about this solution shows that 
the propagator at $\alpha=1/2$ only leads to an exponential damping. Therefore, this non-Markovian treatment is capable of locating the coherent-incoherent transition at the correct spin-boson coupling strength  $\alpha=1/2$. 

\subsubsection{Discretised spin-boson model}

%%%%%%%%%%%%%%%
 \begin{figure}[t]
\centering
\includegraphics[width=.6\columnwidth]{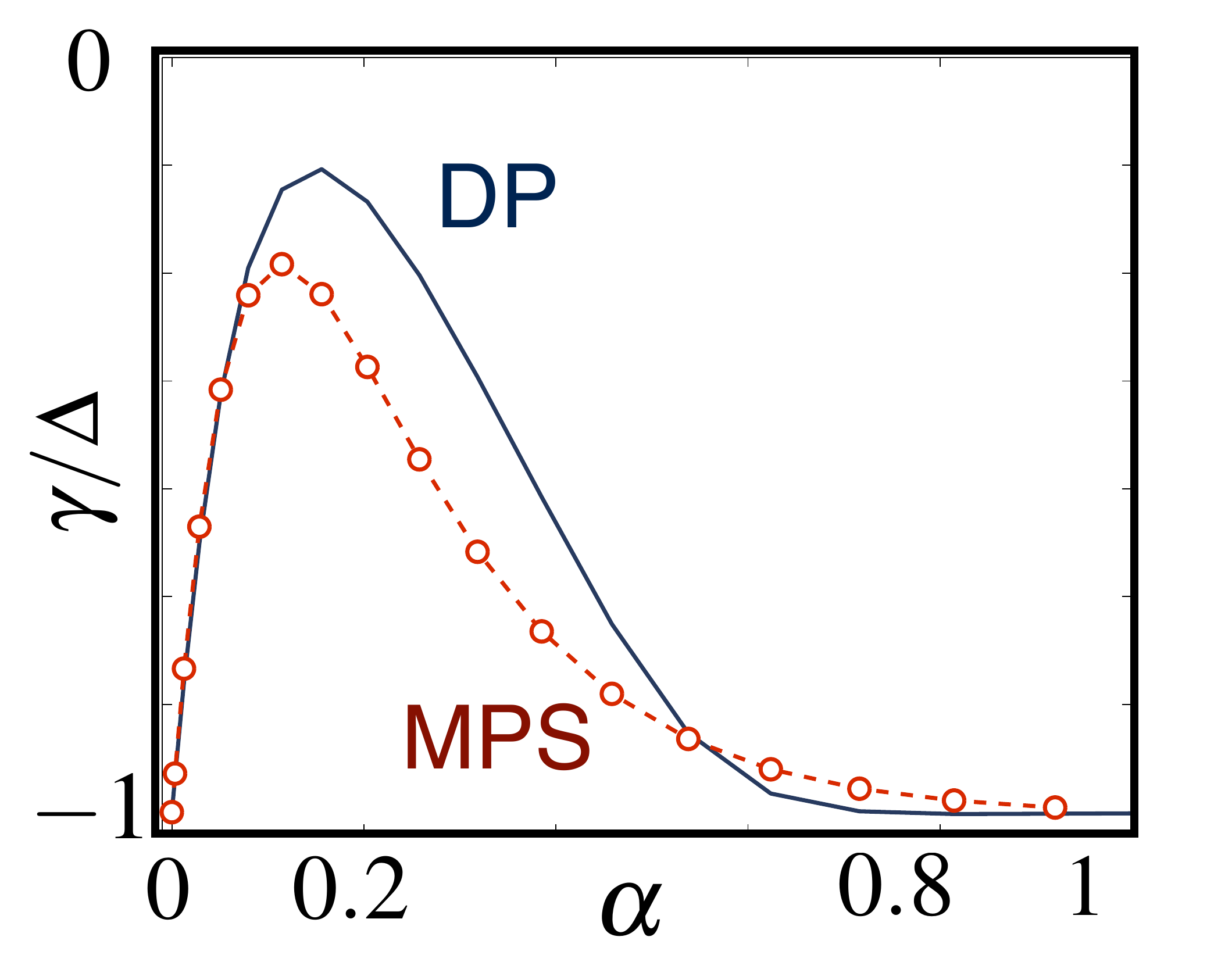} 
\caption{ {\bf(Color online) Spontaneous emission rate:} In blue solid line, calculated with the discrete polaron ansatz for $N=301$ and $L=10\lambda_0$. In dotted red line and for the same input parameters we represent the MPS data.
}
\label{fig_gamma_all}
\end{figure}
%%%%%%%%%%%

Let us now consider the dynamics under the discretised spin-boson model, which  describes the spontaneous emission in the physical transmission line by using numerical means, and can be benchmarked again with MPS simulations, serving thus as a test of the validity of our method. 

Regarding the dynamical polaron ansatz with such discretised model, a clear advantage is that the vector of time-dependent variational parameters  becomes finite 
$\boldsymbol{\alpha}(t)=(\alpha_{1}(t),\alpha_{k_1}(t),\alpha_{k_2}(t),\cdots \alpha_{k_N}(t))^{\rm t}$, and one can directly solve the Schr\"{o}dinger equation $\ii\partial_t\boldsymbol{\alpha}=\mathbb{H}_{\rm P}\boldsymbol{\alpha}$ without making any connections to the Weisskopf-Wigner typical approximations. From a numerical perspective, we can extend this ansatz, at the same computational effort, to consider also the amplitudes of the groundstate in the presence/absence of one additional photonic excitation for the different modes. 
We thus obtain a $2(N+1)\times 2(N+1)$ matrix $\mathbb{H}_{\rm P}$ by evaluating numerically Eq.~\eqref{matrix_elements} for the different matrix elements. Then, the system of differential equations is solved numerically after specifying a particular initial condition, such as $\alpha_1(0)=1$.

In order to study the spontaneous emission rate, we know that the probability amplitude of the qubit in the excited state $\alpha_1(t)$ oscillates with frequency $\Delta_{\rm r}+\delta_1$, and decays with $\gamma_1/2$. If we consider the expectation value of the excitation number operator $\sigma_1^+ \sigma_1^-$, we are able to directly extract the spontaneous emission rate via exponential fitting of 
\begin{equation}
\boldsymbol{\alpha}(t)^\dagger(\sigma_1^+ \sigma_1^-)\boldsymbol{\alpha}(t)=|\alpha_1(t)|^2=\ee^{-\gamma_1 t},
\end{equation}
which we plot as a function of the coupling strength in Fig.\ \ref{fig_gamma_all}, alongside with the same quantity obtained with the MPS ansatz. The agreement of both approaches is remarkable, and serves as a test of the validity of the proposed dynamical polaron ansatz. Let us remark that, by solving directly the variational Schr\"{o}dinger equation, no Markovian approximations are taken with the discretised ansatz, and it should thus give more accurate predictions than the continuum results based on this assumption.

%%%%%%%%%%%%%%%
 \begin{figure}
\includegraphics[width=\linewidth]{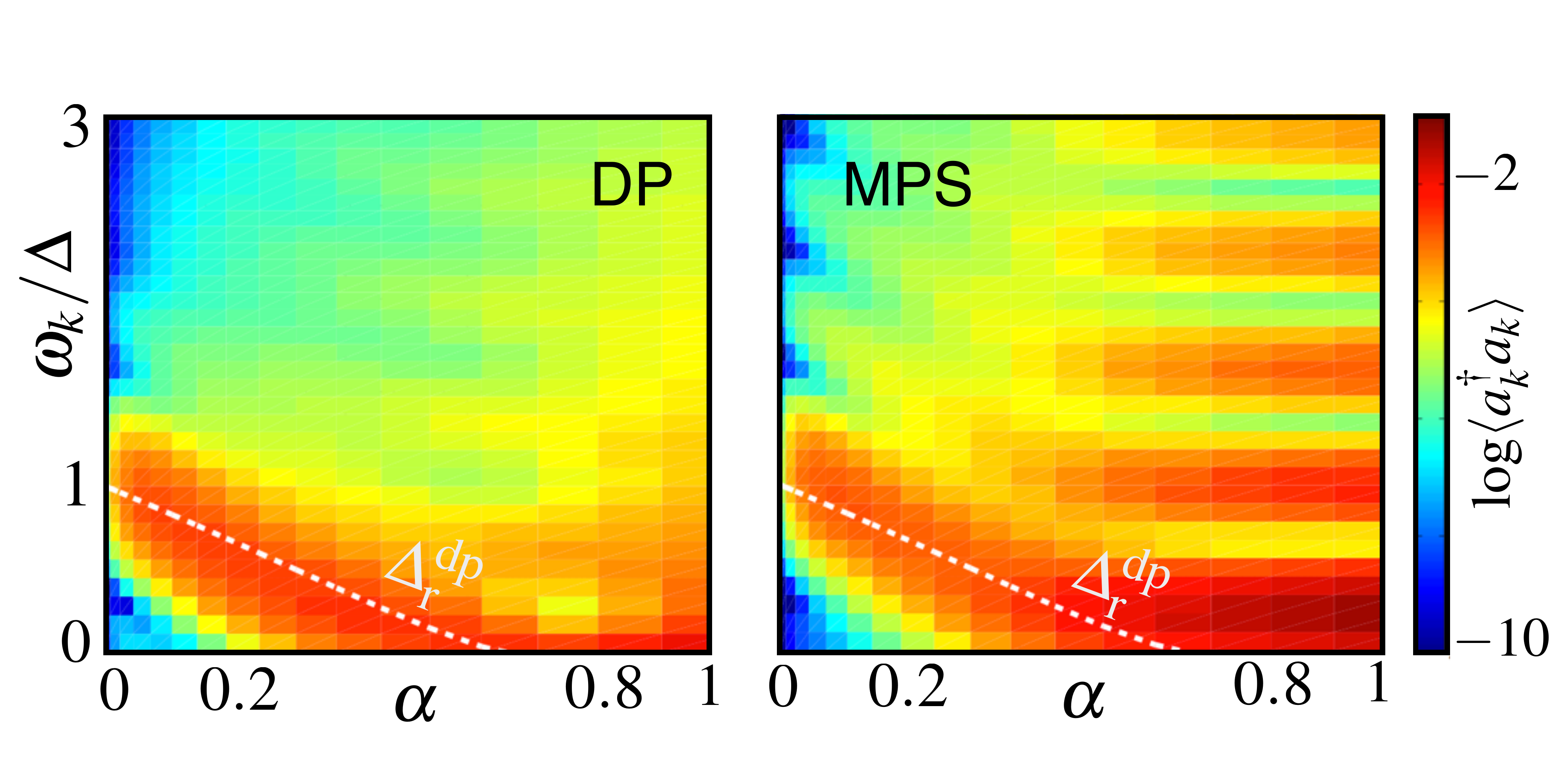} 
\caption{ {\bf (Color online) Photon density in the spontaneous emission:} (left) discrete polaron ansatz, and (right) MPS ansatz. The z-axis in these plots has a logarithmic scale and so does the common colorbar. In white dashed line the renormalized qubit frequency is displayed over the surface plots, confirming visually that the bosonic resonances coincide with it. Beyond the point of $\alpha=0.5$, the two models display some divergences.  
}
\label{fig_emission}
\end{figure}
%%%%%%%%%%%

An important and crucial advantage of the discretised model is that,  for the same effort, one can get information about the dynamics of the photons. As emphasized previously, this is rather unique to our method, and very important in light of the current experiments. For instance, it may be of interest  to study the photon number density $\langle a_k^\dagger a_k^{\phantom{\dagger}}\rangle $  after the qubit has relaxed completely. It will give an indirect measure of the qubit frequency akin to a spectroscopy experiment, since the emitted photon is expected to have the same energy as the qubit excitation. 

 We calculated the photon distribution as a function of the coupling strength and the frequency of the modes within our discrete polaron ansatz,  and compared the predictions to   another simulation with the same parameters using the MPS ansatz. As shown in Fig.~\ref{fig_emission}, both methods show clearly a peak in the distribution around the renormalized qubit frequency calculated through the static ansatz in the previous section, and thus confirm the above intuition that this photonic observable serves as a spectroscopy probe. Let us note that, since we are using periodic boundary conditions for the transmission line,  time must be long enough for the spin excitation to have decayed, but sufficiently short  that  the photon cannot get around the line and scatter with the emitter again (revival time).

%%%%%%%%%%%%%%%
 \begin{figure*}
\centering
\includegraphics[width=1.3\columnwidth]{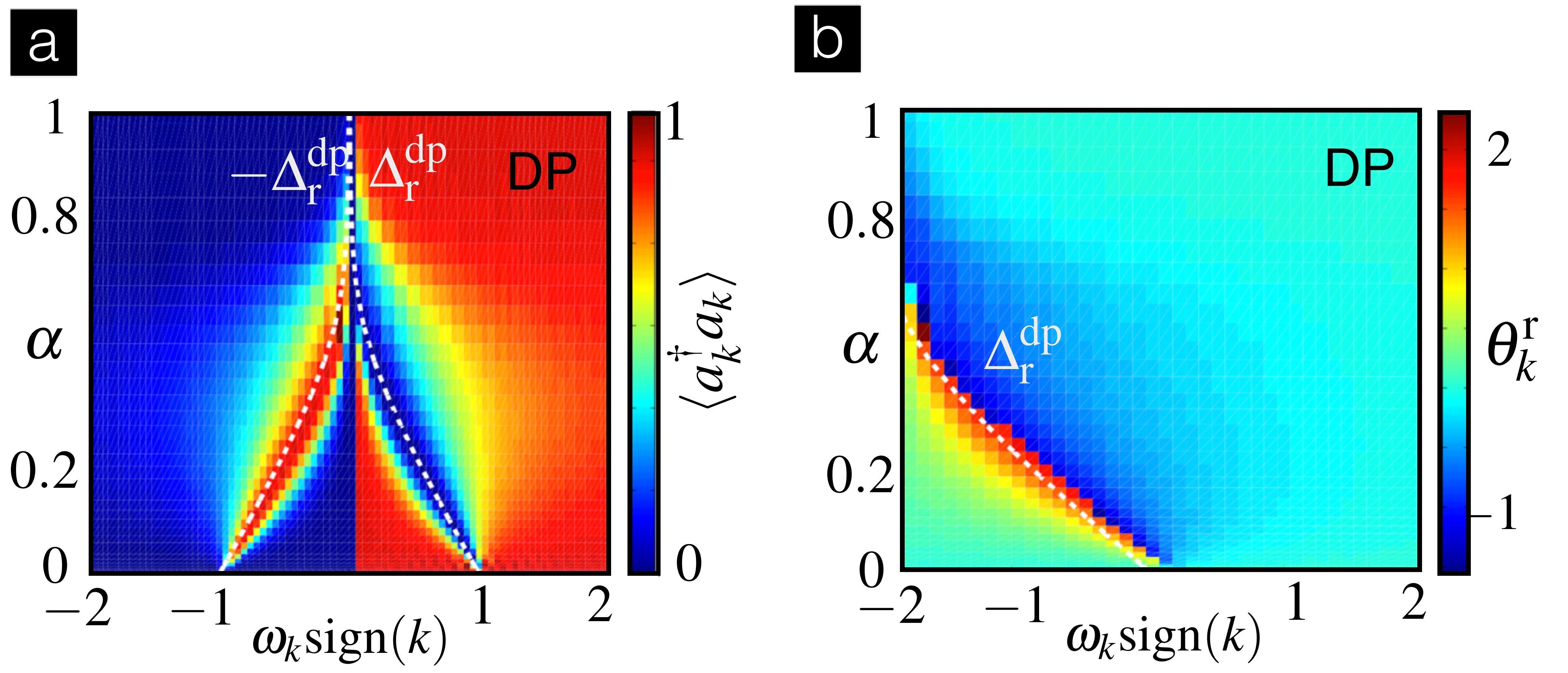} 
\caption{ {\bf (Color online) Normalized photon density in the single photon scattering:} {\bf (a)} The horizontal axis does not display the frequency $\omega$ but the product of $\omega$ and the sign of the momentum, to tell apart modes with the same energy but opposite travelling direction.  White lines represent the renormalized photon frequency, which coincides with the resonances as expected. The right side of the graph corresponds to the norm of the transmission coefficient $|\mathfrak{T}|^2$ and the left side to the norm of the reflection coefficient $|\mathfrak{R}|^2$. {\bf (b)} Phase shift of the reflection coefficient in the single photon scattering simulation.}
\label{fig_scattering_SH}
\end{figure*}
%%%%%%%%%%%

\subsection{Single photon scattering}

The other type of experiment that can be simulated with these tools is the scattering of one photon travelling in the transmission line with one qubit impurity. The photon distribution, after the collision has occurred, will give information about the transmission and reflection coefficients of the photons in the transmission line ($\mathfrak{T}$ and $\mathfrak{R}$ respectively). 
	
In the polaron-transformed frame, we can consider the initial state as a product of its two components, the qubit in the ground state and the photon distribution specified by the probability amplitudes $\alpha_{k_n}(0)$, namely
\begin{equation}
  \label{init_scattering}
  \ket{\psi_{\rm exc}(0)}=\sum_n\alpha_{k_n}(0)a_{k_n}^\dagger\ket{0}\otimes\ket{\psi_{\rm gs}}.
\end{equation}
During the collision, and some time after it has happened, photon and qubit interact such that the wavefunction is no longer in  a product state. However, for long-enough times,  the qubit will have decayed completely, and we can describe the state again as 
\begin{equation}
  \ket{\psi_{\rm exc}(\infty)}=\sum_n\alpha_{k_n}(\infty)a_{k_n}^\dagger\ket{0}\otimes\ket{\psi_{\rm gs}},
\end{equation}	 
with a different photon distribution specified by the probability amplitudes $\alpha_{k_n}(\infty)$.
This enables us to define the transmission and reflection coefficients as
\begin{align}
\label{phase}
\mathfrak{r}_{k_n}&=\frac{|\alpha_{-k_n}(\infty)|}{|\alpha_{+k_n}(0)|}e^{\ii  \theta^{\rm r}_{{k_n}}} e^{\ii \omega_{k_n} t},\\
\mathfrak{t}_{k_n}&=\frac{|\alpha_{+k_n}(\infty)|}{|\alpha_{+k_n}(0)|}e^{\ii  \theta^{\rm t}_{{k_n}}} e^{\ii \omega_{k_n} t},\nonumber
\end{align}
being $\theta^{\rm t}_{{k_n}}$ and $\theta^{\rm r}_{{k_n}}$ the complex phases of the coefficients, and $e^{\ii \omega_{k_n} t}$ include all the phases due to the evolution. The calculation of  transmission and reflection coefficients, $\mathfrak{T}_{k_n}=|\mathfrak{t}_{k_n}|^2$ and $\mathfrak{R}_{k_n}=|\mathfrak{r}_{k_n}|^2$, can be achieved from the expectation value of the photon number operator.
	 
Instead of having an initially excited  qubit, this time we initialise  the system with one single photon travelling in one direction of the line with the qubit in its ground state, which corresponds to Eq.~\eqref{init_scattering} for a particular set of probability amplitudes. This photon will be very localized, and will have a flat frequency distribution, but only in the positive momentum. After the scattering, some of the modes will be absorbed by the qubit and re-emitted afterwards in both directions, effectively being reflected/transmitted from the qubit. This will be depicted by modes with positive momentum vanishing, and some other modes with the same energy but opposite momentum appearing. To check that our polaron predictions for this scattering experiment  are consistent, we  compare the photon density resonances with the renormalized qubit gap obtained through the static ansatz. We can confirm this in Fig.\ \ref{fig_scattering_SH}, which also shows that for high enough couplings, the qubit stops interacting with the photons because its dynamics gets frozen.
	
This analysis of the probability density of the photons  gives the norm of the transmission and reflection coefficients. Additionally, we can also calculate the phase shift as the difference between the final and initial phases, by subtracting the phase due to the time evolution. To perform this task, we run two simulations, the first one with the qubit connected to the line and the second one without it (equivalent to setting the coupling strength to zero). We then extract the phase shift of each mode as the division between the amplitude of that mode in the scattered and in the free case as in Eq.\ \eqref{phase}. Fig.\ \ref{fig_scattering_SH} {\bf (b)} shows this scattering phase for different values of the coupling strength. As expected, the phase jump occurs around the renormalized qubit frequency. 

\section{\bf Conclusions and outlook}
\label{sec:conclusions}

In this work, we have introduced a simple and general  technique  based on a time-dependent variational principle to describe dynamical aspects of a system of quantum emitters ultra-strongly coupled to the 1D EM field  beyond the RWA regime.  The dynamics of the system is described within a region of the Hilbert space that is relevant for the typical experimental situations encountered in spontaneous emission of initially-excited emitters, or scattering  of an incoming photon wave packet by  the quantum emitters. By a physically-motivated parametrisation of this relevant region of the Hilbert space, these dynamical variational ans\"atze become computationally less expensive than Matrix-Product-State simulations, and  allow us to develop some physical insight in certain regimes where analytical results can be derived. More importantly, they allow to address the dynamics of the photonic degrees of freedom, which becomes very relevant in light of the recent experimental progress.

In order to benchmark the accuracy of our variational techniques, we have performed a detailed comparison of static and dynamical predictions for a single quantum emitter ultra-strongly coupled to the EM field. In this context, we have compared the predictions of these simple polaron methods with the numerical results of the quasi-exact MPS methods for both equilibrium and non-equilibrium scenarios, showing a remarkable agreement in typical situations of spontaneous emission and photon scattering. Given the computational simplicity of the introduced ansatz, as compared to the complexity of MPS methods, and its demonstrated reasonable accuracy, we believe that it can become a useful theoretical tool for the quantum optics community interested in the ultra-strong coupling regime. 

The computational advantage of the introduced ansatz becomes even more important as the number of quantum emitters is increased. Although we have focused in the present manuscript on single-emitter applications, the general scheme presented in Sec.~\ref{sec:ansatzs} can be applied to any number of emitters. For instance, we can start from the static polaron ansatz for a couple of quantum emitters $N_{\rm s}=2$\ \cite{two_spin_SH}, and build our dynamical ansatz to analyse  photon-mediated interactions and collective effects in the spontaneous emission. To illustrate some the power of our dynamical ansatz, let us advance some of the predictions that will be detailed elsewhere. 

\subsection{Outlook for two-emitter applications}

The static variational problem for two identical emitters  can be reduced to a couple of implicit equations, one of the renormalised qubit frequencies $\Delta_{\rm r}$, and another one for the photon-mediated Ising interactions $J_{\rm I}$, which display an interesting dependence with the inter-qubit distance\ \cite{two_spin_SH}. Let us discuss two particularly simple limits: {\it (a)} at large distances, the interactions are  vanishingly small,  such that 
 \begin{equation}
\label{large_d}
{\rm J}_{\rm I}\to0,\hspace{2ex}\Delta_{\rm r}=\Delta\left(p_{\infty}\frac{\Delta}{\omega_{\rm c}}\right)^{\frac{\alpha}{1-\alpha}},
\end{equation}
where $p_\infty=p$ coincides with the single-emitter predictions~\eqref{Deltar_continuum_single_qubit}. We thus see that for $\alpha=1$, the  frequencies of both qubits renormalize to zero in the scaling limit, such that the localization-delocalization transition of  both impurities occurs at the same critical coupling as the single-impurity ohmic spin-boson model~\eqref{Deltar_continuum_single_qubit}. {\it (b)} Conversely, at short distances $d\ll v/\omega_{\rm c}$,  the interactions are ferromagnetic and increase with the cutoff
\begin{equation}
\label{short_d}
{\rm J}_{\rm I}\to-\alpha\omega_c,\hspace{2ex}\Delta_{\rm r}=\Delta\left(p_0\frac{\Delta}{\omega_{\rm c}}\right)^{\frac{2\alpha}{1-2\alpha}},
\end{equation}
where $p_0=(p/\alpha)^{1/2}$, provided that $\alpha\omega_{\rm c}/\Delta\gg1$. In this case, the renormalized frequency vanishes at a smaller spin-boson coupling $\alpha=\half$, which was used in\ \cite{two_spin_SH} to locate the localisation-delocalization phase transition.

Turning our attention onto the dynamical effects, let us note that the qubits can also exchange real photons beyond the virtual photons associated to the polaron clouds, and this leads to collective effects in the two-impurity  spin-boson model. In order to account for these effects, we note that the variational spin Hamiltonian~\eqref{Ising_model} evaluated at $[f_{ik}^{\star}]$ leads to two spin-wave excitations that can be excited from the groundstate if the spins absorb a real photon from the EM environment. In the  Markovian limit of the Weisskopf-Wigner-type theory analogous to the single-emitter case~\eqref{markovian_WW_1}, we find that the amplitudes of the excitations fulfil 
\begin{equation}
\begin{split}
\label{markovian_WW_2}
\ii\partial_{t}\alpha_1'(t)&=\left(\delta_1-\ii\frac{\gamma_1}{2}\right)\alpha_1'(t)+\left(g_{12}-\ii\frac{\gamma_{12}}{2}\right)\alpha_2'(t),\\
\ii\partial_{t}\alpha_2'(t)&=\left(\delta_2-\ii\frac{\gamma_2}{2}\right)\alpha_2'(t)+\left(g_{21}-\ii\frac{\gamma_{21}}{2}\right)\alpha_1'(t),\\
\end{split}
\end{equation}
where we have introduced the single-qubit $\gamma_i$ and collective relaxation rates $\gamma_{12}=\gamma_{21}$, as well as the  single-qubit $\delta_i$ Lamb shifts, and collective interactions $g_{12}=g_{21}$. Let us comment on the particular expressions for these parameters in order.

\vspace{.5ex}
{\it (i) Spontaneous decay rates.-} The incoherent spontaneous emission is given by the following  decay rates 
\begin{equation}
\label{i_c_rates}
\begin{split}
\gamma_i=J(\Delta_{\rm r}\zeta)\chi^2,\hspace{3ex} 
\gamma_{12}=\gamma_1\cos\left(\frac{\Delta_{\rm r}\zeta}{v}d\right).
\end{split}
\end{equation}
where we  introduced  $\zeta=(1+{\rm J}_{\rm I}^2/\Delta_{{\rm r}}^2)^{1/2}$, $\chi=\zeta\eta/(1+\zeta^2)$, and $\eta=\{\sqrt{\zeta+1}(\zeta^{-1}+1)+\sqrt{\zeta-1}(\zeta^{-1}-1)\}/\sqrt{2\zeta}$. 

These expressions allow us to study how the spontaneous emission of the pair of qubits 
gets modified as the coupling to the EM field increases, eventually entering the ultra-strong coupling regime. Regarding the individual decay rates, their change with respect to the weak-coupling value is caused by both the renormalization of the qubit frequency $\Delta_{\rm r}<\Delta$, and the Ising interactions due to virtual photon exchange $\zeta>1$. As a consequence, the individual decay rates will also depend on the inter-qubit distance. In the scaling limit $\Delta\ll\omega_{\rm c}$, using the results form the previous section~\eqref{large_d}-\eqref{short_d}, we find
\begin{equation}
\label{decay_rates}
\gamma_i=\left\{\begin{array}{c}\pi\alpha\Delta\left(p_\infty\frac{\Delta}{\omega_c}\right)^{\frac{\alpha}{1-\alpha}}, \hspace{8ex}d\to\infty, \\ \pi\alpha\chi_0^2\zeta_0\Delta\left(p_0\frac{\Delta}{\omega_c}\right)^{\frac{2\alpha}{1-2\alpha}}, \hspace{3ex}d\to0,
\end{array}
\right.
\end{equation}
where we have introduced  $\zeta_0=(1+(\alpha\omega_{\rm c}/\Delta_{\rm r})^2)^{1/2}$, and $\chi_0, \eta_0$ are the above parameters evaluated at $\zeta=\zeta_0$.

Let us first comment on the regime where the qubits are so far apart that the Ising interaction is negligible. When the spin-boson coupling is sufficiently weak $\alpha\ll1$, we recover the result  expected from the usual Weisskopf-Wigner theory of spontaneous emission  $\gamma_i=J(\Delta)=\pi\alpha\Delta$ which rests on  the rotating wave approximation (RWA).
Regarding the collective spontaneous emission in this weak-coupling regime, we observe that it gets suppressed for distances $d=v(2n+1)\pi/2\Delta$ with $n\in\mathbb{Z}^+$, which coincides again with the predictions based on the RWA\ \cite{two_spin_dissipative_entanglement,many_spin_diss_entanglement,rwa_paper}. On the contrary, the collective spontaneous emission attains a maximum at $d=vn\pi/\Delta$ with $n\in\mathbb{Z}^+$, leading to sub/superadiant  channels related to the singlet/triplet Bell states being dark states\ \cite{two_spin_dissipative_entanglement}.

These predictions are modified as the spin-boson coupling is raised, or as the qubits get closer. By raising the spin-photon couplings, still  at large distances, we see that the individual   emission rates depend on the spectral density evaluated at the renormalized qubit frequencies  $\gamma_i=J(\Delta_{\rm r})=\pi\alpha\Delta(p_\infty\Delta/\omega_c)^{\alpha/(1-\alpha)}$. Therefore,  the individual and collective spontaneous emission  get totally suppressed in the localized regime $\alpha=\alpha_{\rm c}^{\infty}=1$ at sufficiently large distances. In the delocalized regime,  $\alpha<\alpha_{\rm c}^{\infty}=1$,  both individual and collective incoherent decay channels contribute to the dynamics. In contrast to the RWA, the collective rates get  suppressed (enhanced) at distances that are also controlled by the renormalized frequency $d=v(2n+1)\pi/2\Delta_{\rm r}$ ($d=vn\pi/2\Delta_{\rm r}$) with $n\in\mathbb{Z}^+$. Therefore, in order to exploit such collective decay to engineer entangled states dissipatively, as proposed in\ \cite{two_spin_dissipative_entanglement,many_spin_diss_entanglement}, it is very important to estimate the correct distance dependence  by a careful calculation of the renormalized frequency.

Let us now move onto the other regime where the standard Weisskopf-Wigner predictions are modified: short inter-qubit distances. Even in the weak-coupling regime where $\Delta_{\rm r}\approx\Delta$, the strength of the individual decay rates,  $\gamma_i=J(\Delta\zeta_0)\chi_0^2=\pi\alpha\Delta(\chi_0^2\zeta_0)$, is different from the RWA predictions $\gamma_i=J(\Delta)=\pi\alpha\Delta$ as a consequence of the ferromagnetic Ising interactions $\chi_0^2\zeta_0\neq 1$. The distance where the collective decay rates get  suppressed (enhanced) is also modified by the presence of interactions $d=v(2n+1)\pi/2\Delta\zeta_0$ ($d=vn\pi/2\Delta\zeta_0$) with $n\in\mathbb{Z}^+$. Hence, in order to exploit these collective decay to engineer entangled states dissipatively, it is also  important to estimate the correct distance by a careful calculation of the interactions due to virtual photon exchange. The differences with respect to the standard Weisskopf-Wigner theory become more important as the spin-boson coupling is increased.  As shown in Eqs.~\eqref{decay_rates} and~\eqref{i_c_rates},  the individual and collective spontaneous emission  are suppressed in the localized regime, which at short distances occurs for a weaker coupling $\alpha=\alpha_{\rm c}^{0}=1/2$.

\vspace{.5ex}
{\it (ii) Lamb shifts and photon-mediated interactions.-}
In addition to the above incoherent decay rates, the qubits will also suffer a frequency shift and a coherent interaction due to the exchange of real photons (i.e. on-shell contribution).  The frequency shifts (i.e. Lamb shifts) of the qubits due to the photonic environment  can be expressed  as
\begin{equation}
\delta_1=\delta_2=-\alpha\frac{\eta\chi}{2}\Delta_{\rm r}\left(1-f_{\rm L}\left(\frac{\Delta_{\rm r}}{\omega_{\rm c}\zeta}\right)\right),
\end{equation}
where $f_{\rm L}(x)=\frac{\zeta^2}{1+\zeta^2}{\rm Re}\left\{\ee^{x}{\rm E}_1(x)-\ee^{-x\zeta^2}{\rm E}_1(-x\zeta^2)\right\}$ is defined  in terms of the exponential integral ${\rm E}_1(z)=\int_{z}^\infty{\rm d}t \ee^{-t}/t$. In the scaling limit, and according to Eqs.~\eqref{large_d}-\eqref{short_d}, the Lamb shifts become
\begin{equation}
\label{Lamb_shifts}
\delta_i=\left\{\begin{array}{c}-\alpha\Delta\left(p_\infty\frac{\Delta}{\omega_c}\right)^{\frac{\alpha}{1-\alpha}}, \hspace{20ex}d\to\infty, \\ -\alpha\frac{\eta_0\chi_0}{2}\left(1-\frac{\zeta_0^2\log\zeta_0^2}{1+\zeta_0^2}\right)\Delta\left(p_0\frac{\Delta}{\omega_c}\right)^{\frac{2\alpha}{1-2\alpha}}, \hspace{2ex}d\to0,
\end{array}
\right.
\end{equation}
As occurs for the single-emitter case, a  calculation beyond the Markovian approximation should be performed to locate exactly the coherent-incoherent transition.

Let us now focus on  the coherent photon-mediated interactions $g_{12}$, which have two contributions. As argued in the previous section, ${\rm J}_{\rm I}$ is caused by the exchange of virtual off-shell photons. The remaining contribution corresponds to the exchange of real on-shell photons. In particular, again in the scaling limit, we find
\begin{equation}
\label{g12}
g_{12}={\rm J}_{\rm I}+\frac{\pi}{2}\zeta\chi^2\alpha\Delta_{\rm r}\sin\left(\frac{\Delta_{\rm r}\zeta }{v}d\right)+\delta g_{12}\left(\ii\frac{\Delta_{\rm r} }{v\zeta}d+\frac{\Delta_{\rm r}}{\zeta\omega_{\rm c}}\right),
\end{equation}
where we have introduced 
\begin{align}
\delta g_{12}(z)&=\frac{-\chi^2\alpha\Delta_{\rm r}}{2}
\bigg(\!1+\!{\rm Im}zf_{\rm I}(z)-\zeta\!\left(f_{\rm R}(z)-f_{\rm R}\big(-z^*\zeta^2\big)\!\right)\!\!\bigg),\nonumber
\end{align}
and we have used  the following  functions 
\begin{equation}
\label{functions}
f_{\rm I}(z)={\rm Im}\left\{{\rm E}_1(z)\ee^{z}\right\},\hspace{2ex} f_{\rm R}(z)={\rm Re}\left\{{\rm E}_1(z)\ee^{z}\right\}.
\end{equation}

At large distances, where the Ising contribution $J_{\rm I}$ due to virtual photon exchange vanishes~\eqref{large_d}, all the qubit-qubit interaction is due to the exchange of real photons. Moreover, $\delta g_{12}$ also vanishes at large distances, and we obtain
\begin{equation}
\label{large_d_g12}
g_{12}=\frac{\pi}{2}\alpha\Delta_{\rm r}\sin\left(\frac{\Delta_{\rm r} }{v}d\right).
\end{equation}
Let us first consider the weak-coupling limit $\alpha\ll 1$, where the qubit  frequencies approach the bare value $\Delta_{\rm r}\approx\Delta$. In this case, we obtain $g_{12}=\frac{\pi}{2}\alpha\Delta\sin\left(\frac{\Delta }{v}d\right)$, in accordance with the results based on the RWA\ \cite{rwa_paper}. We thus recover the result that the photon-exchange interactions are suppressed for the distances $d=vn\pi\Delta$ with $n\in\mathbb{Z}^+$, where the collective spontaneous decay~\eqref{i_c_rates} is maximal, and viceversa\ \cite{two_spin_dissipative_entanglement}. As the spin-boson coupling is increased, these distances are changed as a consequence of the renormalization of the qubits frequency $d=vn\pi\Delta_{\rm r}$ with $n\in\mathbb{Z}^+$. However, when the spin-photon coupling is sufficiently large, the interactions are totally suppressed since
\begin{equation}
|g_{12}|\leq \frac{\pi}{2}\alpha\Delta\left(p_\infty\frac{\Delta}{\omega_c}\right)^{\frac{\alpha}{1-\alpha}}\to 0,\hspace{2ex} \alpha\to \alpha_{\rm c}^\infty=1,
\end{equation}
and no coherent swap of the excitation can occur. We thus see that in the localized phase, all coherent and incoherent processes are inhibited.

The situation changes considerably at very short distances, where  the coherent Ising part $J_{\rm I}$ given by Eq.~\eqref{short_d} becomes the leading term in the scaling limit. In this case, the interactions
\begin{equation}
\label{short_d_g12}
g_{12}=-\alpha\omega_{\rm c}-\half\alpha\chi_0^2\Delta_{\rm r}\approx-\alpha\omega_{\rm c},
\end{equation}
become independent of the renormalized qubit frequency, and diverge with the cutoff frequency.  Such behaviour is  consistent with the results based on the RWA approximation,
and predict that the excitation can always be coherently swapped between the  qubits, provided that they are close enough. In the localized regime, $ \alpha\geq\alpha_{\rm c}^0=\half$, the decay channels are suppressed, such that the qubits  continue swapping the excitation indefinitely.

Let us finally highlight that, just as the simple Markovian approximation leading to Eqs.~\eqref{markovian_WW_2} can be improved to predict the coherent-incoherent transition by taking into account 
non-Markovian effects, it can also be improved by taking into account retardation times for the exchange of photons\ \cite{milonni_knight_causality}, thus making an interesting connection with the emerging causality for spin-boson models discretised on a lattice\ \cite{LRB_spin_boson}.

\acknowledgments
The authors acknowledge support from Spanish Mineco Project FIS2012-33022, CAM Research Network QUITEMAD+ and EU FP7 FET-Open project PROMISCE.

\end{document}